\documentclass[amsmath,amssymb, aps, pra,reprint,superscriptaddress]{revtex4-1}

\usepackage{float}
\usepackage{amsmath,amssymb}
\usepackage[english]{babel}
\usepackage{ bbold }
\usepackage{upgreek}
\usepackage{dsfont}
\usepackage{graphicx}
\usepackage{graphicx}

\newcommand{\op}[1]{{\sf #1}}

\newcommand{\cD}{{\mathcal D}}
\newcommand{\cE}{{\mathcal E}}
\newcommand{\cL}{{\mathcal L}}

\newcommand{\oH}{{\mathsf H}}

\newcommand{\ket}[1]{\left \vert #1 \right \rangle}

\newcommand{\matel}[3]{{\left\langle \vphantom{#1 #2 #3} #1 \,\right\vert
\left.
 \hspace{-0.15em} \vphantom{#1 #2 #3} #2 \,\right\vert \left.
 \hspace{-0.15em} \vphantom{#1 #2 #3} #3\right\rangle}}
\newcommand{\ketbra}[2]{\vert #1 \rangle \langle #2 \vert}
\newcommand{\be}{\begin{equation}}
\newcommand{\ee}{\end{equation}}
\newcommand{\bea}{\begin{eqnarray}}
\newcommand{\eea}{\end{eqnarray}}

\newcommand{\un}{{\mathds 1}}

\newcommand{\oJ}{\textsf{\textbf{J}}}
\newcommand{\orcm}{\textsf{\textbf{r}}}
\newcommand{\om}{\textsf{\textbf{m}}}

\newcommand{\oP}{\textsf{\textbf{P}}}
\newcommand{\oX}{\textsf{\textbf{X}}}

\newcommand{\oL}{\op{L}}
\newcommand{\of}{\op{f}}
\newcommand{\oq}{\textsf{\textbf{q}}}

\newcommand{\rB}{\mathrm{B}}
\newcommand{\rf}{\mathrm{f}}
\newcommand{\rV}{\mathrm{V}}

\begin{document}

\title{Rotational Alignment Decay and Decoherence of Molecular Superrotors}

\author{Benjamin A. Stickler}
\affiliation{University of Duisburg-Essen, Faculty of Physics, Lotharstra\ss e 1, 47048 Duisburg, Germany}

\author{Farhad T. Ghahramani}
\affiliation{School of Physics, Institute for Research in Fundamental Sciences (IPM), P.O. Box 19395-5531, Tehran, Iran}

\author{Klaus Hornberger}
\affiliation{University of Duisburg-Essen, Faculty of Physics, Lotharstra\ss e 1, 47048 Duisburg, Germany}

\begin{abstract}
We present the quantum master equation describing the coherent and incoherent dynamics of a rapidly rotating molecule in presence of a thermal background gas. The master equation relates the rate of rotational alignment decay and decoherence to the microscopic scattering amplitudes, which we calculate for anisotropic van der Waals scattering. For large rotational energies, we find excellent agreement of the resulting alignment decay rate with recent superrotor experiments.
\end{abstract}

\maketitle

\section{Introduction}
The precise control of molecular rotation dynamics is a challenging task holding the prospect of orientation-resolved metrology, enhancement of chemical reaction rates, and state-selective collision studies \cite{stapelfeldt2003,lemeshko2013,moses2017}. Over the past decades, various techniques were conceived and established to manipulate the orientation and rotation state of small molecules and nanoparticles \cite{stapelfeldt2003,ohshima2010,fleischer2012,lemeshko2013,moses2017}. Recently, it was demonstrated that optical centrifuge beams \cite{karczmarek1999} can produce rotational wavepackets of unprecedentedly high angular momentum \cite{villeneuve2000,yuan2011,toro2013,milner2014a,korobenko2014,milner2015,milner2017}. The high rotation rates of such {\it molecular superrotors} \cite{li2000} were predicted to suppress the rotational-translational energy transfer \cite{hartmann2012}, so that their center-of-mass motion can be trapped and buffer-gas cooled without appreciably affecting the rotational population \cite{forrey2001,alqady2011}.

This stability of superrotor rotations with respect to collisions was indeed observed experimentally with nitrogen and oxygen superrotors at ambient conditions \cite{milner2014a,milner2014b,milner2015,milner2017}. Classical simulations of a gas of superrotors show that, while the high rotation rates are conserved initially, they rapidly relax towards thermal equilibrium once the rotational-translational energy exchange becomes relevant \cite{khodorkovsky2015,milner2015b}. With a duration of a few nanoseconds, the initial {\it gyroscopic stage} is relatively short in state-of-the-art experiments \cite{milner2014a}, but is expected to be orders of magnitude longer in high vacuum. Notwithstanding its fundamental relevance and significance for future applications, a microscopic quantum theory that quantitatively describes the dynamics of a superrotor in its thermal environment and that predicts the ensuing alignment decay and decoherence timescales is still lacking.

Here we establish the Markovian quantum master equation of a rapidly rotating molecule immersed in a thermal gas. Based on the monitoring approach \cite{hornberger2007,hornberger2008,smirne2010}, this Lindblad-type equation describes, in terms of the exact scattering amplitudes, how the rotation state of a superrotor loses its alignment and decoheres. The master equation preserves the rotational energy, reflecting the fact the collisions are approximately elastic due to the high rotation rates of superrotors. This is in contrast to orientational decoherence of slowly rotating particles \cite{stickler2016b,zhong2016,papendell2017}, applicable if the molecular orientation barely changes during the scattering process, and to quantum rotational thermalization \cite{stickler2018a}, describing how the gas induces linear friction and diffusion of the quantized angular momentum vector.

We calculate the rate of alignment decay for nitrogen superrotors and find remarkable agreement with experimental data \cite{milner2014a}. This demonstrates the predictive power of the theory, which includes no free parameters. The decay rate exhibits a markedly different scaling for large rotational energies than the energy corrected sudden approximation \cite{depristo1979,milner2014a,milner2014b}. The presented master equation is expected to be instrumental for future sensing and metrology applications with superrotors, whose sensitivity will be ultimately limited by the collisional interaction between the rotor and its thermal environment.

\begin{figure}
 \centering
 \includegraphics[width = 0.40 \textwidth]{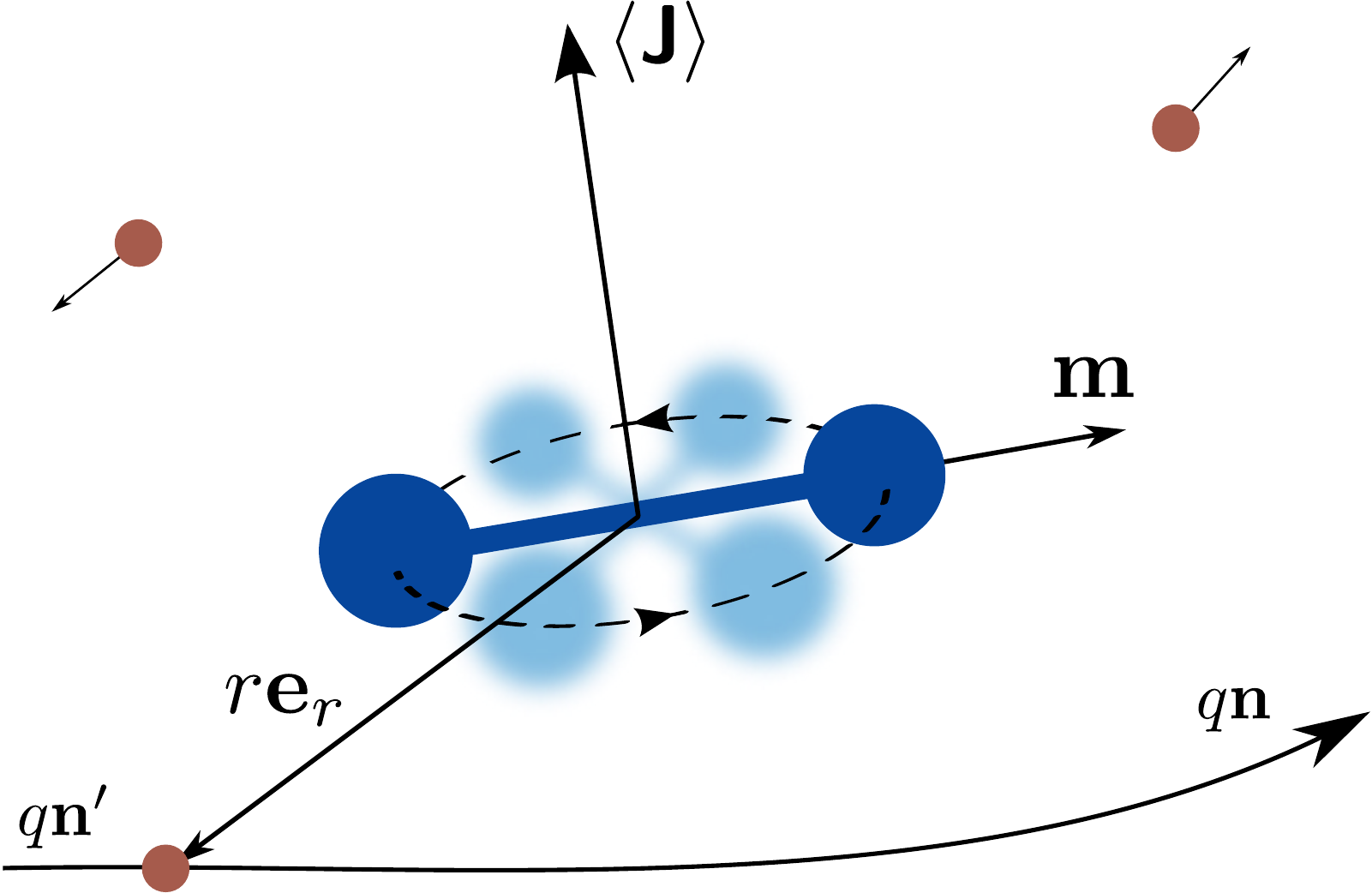}
 \caption{A rapidly rotating linear molecule (blue) embedded in a homogeneous thermal gas (red) of density $n_{\rm g}$ and temperature $T$ experiences collisions with environmental gas particles. If the molecule rotates multiple times during the collision, as in the case of superrotors, the interaction with the gas leads to rotational decoherence as well as re-orientation of the mean angular momentum $\langle {\oJ} \rangle$. The rotor of orientation ${\bf m}$ interacts via the attractive van der Waals interaction \eqref{eq:vdw} with a gas particle at distance $r$, which impinges with relative momentum $q {\bf n}'$ and leaves with relative momentum $q {\bf n}$.} \label{fig:fig1}
\end{figure}

\section{Master Equation}

The dynamics of rotational alignment decay and decoherence can be derived from the monitoring master equation for a massive particle with internal degrees of freedom \cite{smirne2010}. The latter provides a non-perturbative Markovian quantum master equation for a tracer particle moving and revolving in a thermal environment. The description is based on the microscopic collision rate operator and the scattering matrix of an individual collision between the tracer and a gas particle.

In the present case, we consider a linear rigid rotor of mass $M$ and moment of inertia $I$ revolving in a homogeneous monoatomic gas of density $n_{\rm g}$, temperature $T$ and pressure $p_{\rm g} = n_{\rm g} k_{\rm B} T$. For a rapidly rotating superrotor the scattering rate operator as well as the scattering matrix are approximately diagonal in the total angular momentum since the rotor revolves multiple times during the collision with a gas particle (see Fig.~\ref{fig:fig2} and App.~\ref{app:b}). Tracing out the latter together with the thermalized center-of-mass motion yields the master equation $\partial_t \rho = -i \left [ \oH + \oH_{\rm g}, \rho \right ] / \hbar + \cD \rho$ for the rotation state $\rho$  (see App.~\ref{app:a}). The free linear-top Hamiltonian is denoted by $\oH$. The gas affects the rotor dynamics coherently through the energy shift
\begin{subequations}\label{eq:cohincoh}
\begin{equation} \label{eq:coh}
\oH_{\rm g} = - 2 \pi \hbar^2 \frac{n_{\rm g}}{\mu} \int_0^\infty d q q^2 \nu_{\rm th}(q) \int_{S_2} d^2 {\bf n}  {\rm Re} \left [ \of \left ( q{\bf n}, q{\bf n} \right ) \right ]
\end{equation}
and incoherently through the Lindblad superoperator
\begin{align} \label{eq:incoh}
& \cD \rho = \frac{n_{\rm g}}{\mu} \int_0^\infty dq q^3  \nu_{\rm th}(q) \int_{S_2} d^2 {\bf n} d^2 {\bf n}' \notag \\
& \times \left [\of(q {\bf n},q{\bf n}') \rho \of^\dagger(q {\bf n},q{\bf n}')  - \frac{1}{2} \left \{ \of^\dagger(q {\bf n},q{\bf n}')\of(q {\bf n},q{\bf n}'), \rho \right \} \right ].
\end{align}
\end{subequations}

Both contributions \eqref{eq:coh} and \eqref{eq:incoh} are characterized by the thermal distribution of relative momenta $\nu_{\rm th}(q) = \exp ( - q^2/q^2_{\rm th})/(\sqrt{\pi} q_{\rm th})^3$, with the momentum $q_{\rm th} = \sqrt{2 \mu k_{\rm B} T}$ determined by the reduced mass $\mu = m_{\rm g} M/(m_{\rm g} + M)$. In addition, the Lindblad operators are given by the operator-valued microscopic scattering amplitudes
\begin{equation} \label{eq:scattamp}
\of(q {\bf n},q {\bf n}') = \sum_{j = 0}^\infty \sum_{m,m' = -j}^j f_{jm,jm'}(q {\bf n}, q {\bf n}') \ketbra{j m}{jm'},
\end{equation}
for incoming and outgoing relative momentum $q {\bf n}'$ and $q {\bf n}$. Here $\ket{jm}$ is the linear rotor eigenstate with energy $E_j = \hbar^2 j(j+1)/2I$ and azimuthal quantum number $m$. Note that, while Eq.~\eqref{eq:scattamp} conserves the total angular momentum quantum number $j$ of a rapidly rotating molecule, it accounts for elastic scattering to different $m$.

The incoherent term \eqref{eq:incoh} describes how a superrotor loses coherence in a thermal monoatomic gas, while the coherent term \eqref{eq:coh} renormalizes the rotor energy \cite{schmidt2015}. In case that the gas particles have internal degrees of freedom, the scattering amplitudes depend on the incoming and outgoing internal gas state. Tracing out the latter yields the generalization of \eqref{eq:cohincoh} after a straight-forward calculation. We will show next that \eqref{eq:incoh} describes how the quantum state of a rapidly rotating particle re-orients and decoheres due to collisions with surrounding gas particles.

\section{Rotational Decoherence and Alignment Decay}

An optical centrifuge consists of two superimposed laser beams of opposite circular polarization, which  propagate in the same direction along the $z$-axis and whose frequency detuning increases linearly with time \cite{karczmarek1999}. The resulting total electromagnetic field is linearly polarized; the polarization direction rotates with increasing frequency in the $xy$-plane orthogonal to the beam axis. An initially orientationally trapped linear molecule adiabatically follows the field polarization until the laser is instantaneously switched off, releasing the molecules in a superposition of high angular momentum eigenstates $\ket{jj}$. The total state operator $\rho$ of the propelled molecule is thus of the form $\rho = \sum_{j,j' =0}^\infty \rho_{jj'} \ketbra{jj}{j'j'}$, where the coefficients $\rho_{jj'}$ depend on the details of the centrifuge beam \cite{armon2016,armon2017}.

On a short timescale, the derived master equation describes how the matrix elements of the superrotor state $\rho$ oscillate and decay, i.e. $\partial_t \rho_{jj'} \simeq  - (i \Delta_{jj'} + \gamma_{jj'} ) \rho_{jj'}$, with the decay rate
\begin{align} \label{eq:rate}
& \gamma_{jj'} = \frac{n_{\rm g}}{2 \mu} \int_0^\infty dq q^3 \nu_{\rm th}(q) \int_{S_2} d^2{\bf n} d^2{\bf n}' \notag \\ 
 & \times \left [ \Bigl \vert f_{jj,jj}(q {\bf n},q {\bf n}') - f_{j'j',j'j'}(q {\bf n},q {\bf n}') \Bigr  \vert^2 \vphantom{\sum_{m' = -j'}^{j'-1}} \right. \notag \\
 & \left. + \sum_{m = -j}^{j-1} \left \vert f_{jm,jj}(q {\bf n},q {\bf n}') \right \vert^2  + \sum_{m' = -j'}^{j'-1} \left \vert f_{j'm',j'j'}(q {\bf n},q {\bf n}') \right \vert^2 \right ].
\end{align}
The first term describes decoherence between the states $\ket{jj}$ and $\ket{j'j'}$ due to the information acquired by the scatterer during the collision \cite{hornberger2007}. While this first term vanishes for $j = j'$, the remaining two terms are always non-zero since they describe how the plane of rotation of the superrotor changes due to collision-induced transitions, $jj \to jm$ with $m= -j,\ldots,j-1$. This re-orientation of the molecular angular momentum vector is the dominant contribution to the experimentally observed decay of rotational coherences of nitrogen superrotors \cite{milner2014a,milner2014b}, as will be demonstrated below.

We remark that isotropic states, $\sum_{j = 0}^\infty\rho_j  \sum_{m = -j}^j \ketbra{jm}{jm}$, are stationary under the superoperator $\cD$ irrespective of their distribution $\rho_j$ of rotational energies. An arbitrary initial superrotor state with matrix elements $\rho_{jj'}$ will thus approach the isotropic, mixed state with energy distribution $\rho_{jj}$ on the timescale $1/\gamma_{jj'}$, provided that scattering to all $m$ is allowed. The master equation therefore predicts the timescale on which ensembles of superrotors can be used for state-selective collision studies.

\section{van der Waals scattering}

Comparing the decoherence and alignment decay rate \eqref{eq:rate} to experiments requires solving the scattering problem of an individual collision between the superrotor and a gas particle. As a generic situation, we consider scattering of gas atoms off a diatomic molecule interacting via the attractive anisotropic van der Waals potential \cite{hirschfelder1964}
\begin{equation} \label{eq:vdw}
V(r {\bf e}_r, {\bf m}) = - \frac{C_6}{r^6} \left [ 1 + \frac{2 \Delta \alpha}{3 \overline{\alpha}} P_2 ( {\bf e}_r \cdot {\bf m} ) \right ].
\end{equation}
Here, $C_6$ quantifies the strength of the van der Waals interaction and $\Delta \alpha = \alpha_\| - \alpha_\bot$ is the polarizability anisotropy of the linear rotor with $\alpha_\|$ and  $\alpha_\bot$ its polarizabilities along the symmetry axis and orthogonal to it; $\overline{\alpha} = ( 2 \alpha_\bot + \alpha_\|)/3$ is the mean polarizability. The relative distance vector from the molecular center-of-mass to the atom is denoted by $r {\bf e}_r$, ${\bf m}$ is the direction of the rotor symmetry axis, see Fig.~\ref{fig:fig1}, and $P_2(x) = (3 x^2 -1)/2$ is the second order Legendre polynomial.

\begin{figure}
 \centering
 \includegraphics[width = 0.45 \textwidth]{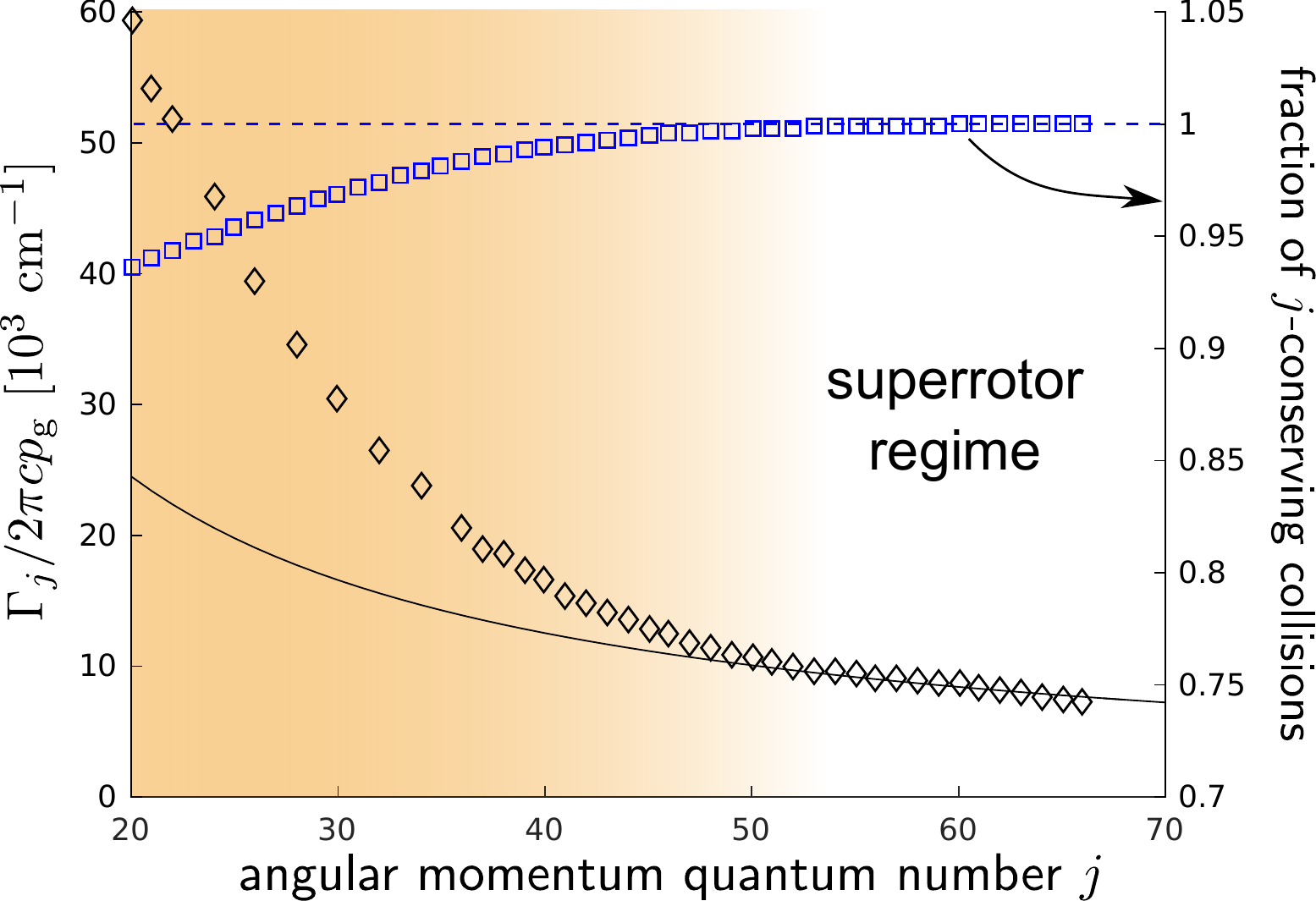}
\caption{In the superrotor regime, characterized by the absence of inelastic collisions, the theoretically predicted decay of rotational coherences (black solid line) compares very well with the experimental data  (black diamonds) \cite{milner2014a}. The theoretical rate \eqref{eq:rate2} includes no free parameters but is based on the microscopic scattering amplitudes of individual superrotor-gas particle collisions. For the considered nitrogen molecules, the superrotor regime starts at $j \simeq 50$, as can be seen from the fraction of $j$-conserving collisions (blue squares, right axis) approaching unity. The elastic fraction is obtained by numerically calculating the total scattering cross sections (using {\sc Molscat} \cite{molscat}) for atom-linear-rotor scattering with a Lennard-Jones potential exhibiting the orientation dependence of Eq.~\eqref{eq:vdw}. The molecular polarizabilities are taken from Ref.~\cite{spelsberg1994} and the $C_6$ constant from the Lennard-Jones parameters listed in Ref.~\cite{mortimer1993}.}\label{fig:fig2}
\end{figure}

In order to calculate the rotation-state-dependent scattering amplitudes $f_{jm,jm'}(q {\bf n},q{\bf n}')$ we use Schiff's approximation \cite{schiff1956} in the limit of a rapidly rotating particle and assume that small-angle scattering dominates the amplitude. After a direct but non-trivial calculation (see App.~\ref{app:b}) one obtains the rates
\begin{subequations} \label{eq:rate2}
\begin{eqnarray}
\gamma_{jj'} &\simeq& \Gamma \left ( \frac{13}{5} \right ) \Gamma^2 \left ( \frac{3}{5} \right )\frac{\sqrt{\pi}n_{\rm g} q_{\rm th}^3}{10 \mu \hbar^2} \left ( \frac{\Delta \alpha}{30 \overline{\alpha}} \right )^2 \notag \\
&& \times \left ( \frac{3 \pi \mu C_6}{8 \hbar q_{\rm th}} \right )^{4/5} A_{jj'}
\end{eqnarray}
whose dependence on the rotation state is described by
\begin{eqnarray}
A_{jj'}& = & \left [ P_2 \left (\frac{2 j}{2 j+1} \right ) - P_2 \left (\frac{2 j'}{2 j'+1} \right )\right ]^2 \notag \\
 && + \frac{1}{6} \left [ P_2^1 \left (\frac{2 j -1}{2 j +1} \right ) \right ]^2 + \frac{1}{6}\left [ P_2^1 \left (\frac{2 j' -1}{2 j' +1} \right ) \right ]^2 \notag \\
 && + \frac{1}{24} \left [ P_2^2 \left (\frac{2 j -2}{2 j +1} \right ) \right ]^2 + \frac{1}{24}\left [ P_2^2 \left (\frac{2 j' -2}{2 j' +1} \right ) \right ]^2.
\end{eqnarray}
\end{subequations}
Here, $P_2^m(x)$ are associated Legendre polynomials.

The decay rate \eqref{eq:rate2} can be directly compared to the decay of rotational coherences observed in Ref.~\cite{milner2014a} with nitrogen superrotors. In this experiment, their alignment decay was monitored by circularly polarized Raman scattering. The resulting signal is proportional to $\vert \matel{jj}{\rho}{j-2j-2} \vert^2$ so that its decay rate $\Gamma_j = 2 \gamma_{jj-2}$ follows directly from Eq.~\eqref{eq:rate2}.

In Fig.~\ref{fig:fig2} we show a comparison between the theoretically expected signal decay rate and the experimentally observed rates as reported in Ref.~\cite{milner2014a}. In the superrotor regime $j \gtrsim 50$, where the rotational energy is approximately conserved during the scattering process, the agreement between theory and experiment is excellent. We emphasize that the theory includes not a single free parameter but is based on the microscopic scattering amplitudes, testifying the predictive power of the approach. For large $j$, the signal decays as $1/j$ due to the asymptotic behavior of the associated Legendre polynomials,  $A_{jj-2} \sim 6/j$ for $j \to \infty$. This is in marked contrast to the prediction $\Gamma_j \sim 1/j^4$ of the energy-corrected sudden approximation \cite{depristo1979,milner2014a}, valid only for weakly non-adiabatic collisions.

\section{Conclusion}

In summary, we established the quantum master equation of a molecular superrotor revolving in a thermal environment. It is based on the scattering amplitudes of superrotor-gas collisions, following directly from the intermolecular interaction potential. The ensuing alignment decay rates agree very well with experimental observations. The theory predicts the timescales for quantum coherent experiments with superrotors. Combined with ab-initio scattering calculations, it can be applied to any superrotating system, paving the way for future state-resolved collision studies and coherence experiments.

\acknowledgments{We thank Valery Milner for sending us the experimental data of Ref.~\cite{milner2014a} and we are grateful for the computational resources provided by the High Performance Computing Center, School of Computer Science, IPM, for the {\sc Molscat} calculations.}

\clearpage
\appendix
\begin{widetext}
 \section{Master Equation}  \label{app:a}

To derive the quantum dynamics of rotational alignment decay and decoherence of molecular superrotors we start from the master equation for a linear molecule of mass $M$ and moment of inertia $I$ immersed in a homogeneous gas of density $n_{\rm g}$, mass $m_{\rm g}$ and temperature $T$. Adapting the theory presented in Ref.~\cite{smirne2010}, one obtains for the combined rotational-center-of-mass state $\rho_{\rm tot}$ the Markovian master equation
\begin{equation}
\partial_t \rho_{\rm tot} = - \frac{i}{\hbar} \left [ \oH_{\rm tot} + \oH_n, \rho_{\rm tot} \right ] + \cL \rho_{\rm tot}.
\end{equation}
Here the total free Hamiltonian is  $\oH_{\rm tot} = \oP^2 / 2 M + \oJ^2/ 2I$, the gas-induced energy shift
\begin{equation}
\oH_n = - 2 \pi \hbar^2 \frac{n_{\rm g}}{\mu} \sum_{j = 0}^\infty \sum_{m,m' = - \ell}^\ell \int d^3 {\bf p} \mu_{\rm g}({\bf p}) {\rm Re} \left [f_{j m, j m'}( {\rm rel}({\bf p},\oP),{\rm rel}({\bf p},\oP) ) \right ] \ketbra{jm}{jm'},
\end{equation}
and the Lindblad superoperator
\begin{equation}
\cL \rho_{\rm tot} = \sum_{\cE} \int d^3 {\bf Q} \int_{{\bf p} \bot {\bf Q}} d^2 {\bf p} \left [ \oL_{{\bf Q} {\bf p} \cE} \rho_{\rm tot} \oL_{{\bf Q} {\bf p} \cE}^\dagger - \frac{1}{2} \left \{\oL_{{\bf Q} {\bf p} \cE}^\dagger \oL_{{\bf Q} {\bf p} \cE}, \rho_{\rm tot} \right \} \right ].
\end{equation}
with the Lindblad operators
\begin{equation}
\oL_{{\bf Q} {\bf p} \cE} = e^{i {\bf Q} \cdot \oX / \hbar} \sum_{\stackrel{jj' = 0}{E_j - E_{j'} = \cE}}^\infty \sum_{m = -j}^j \sum_{m' = -j'}^{j'}L_{jm,j'm'}({\bf p},\oP;{\bf Q}) \ketbra{jm}{j'm'}
\end{equation}
where
\begin{eqnarray}
L_{jm,j'm'}({\bf p},{\bf P};{\bf Q})& =& \sqrt{\frac{n_{\rm g} m_{\rm g}}{\mu^2 Q}} \sqrt{\mu_{\rm g} \left [ {\bf p}_\bot + \frac{m_{\rm g}}{M} {\bf P}_\| + \left (1 + \frac{m_{\rm g}}{M} \right ) \frac{{\bf Q}}{2} + \frac{E_j - E_{j'}}{Q^2/m_{\rm g}} {\bf Q} \right ]} \notag \\
 && \times f_{jm,j'm'} \left [ {\rm rel}({\bf p}_\bot, {\bf P}_\bot) - \frac{{\bf Q}}{2} + \frac{E_j - E_{j'}}{Q^2/\mu} {\bf Q}, {\rm rel}({\bf p}_\bot, {\bf P}_\bot) + \frac{{\bf Q}}{2} + \frac{E_j - E_{j'}}{Q^2/\mu} {\bf Q} \right ].
\end{eqnarray}
In addition to the notation of the main manuscript, $\mu_{\rm g}({\bf p})$ denotes the thermal momentum distribution of the gas, $\oP$ is the center-of-mass momentum operator of the rotor, $\oX$ its position operator, $E_j = \hbar^2 j (j+1)/2 I$ the rotation energy and we defined the relative momentum
\begin{equation}
{\rm rel}({\bf p},{\bf P}) = \frac{\mu}{m_{\rm g}} {\bf p} - \frac{\mu}{M} {\bf P}.
\end{equation}

For superrotors generated by an optical centrifuge the center-of-mass and rotation state are uncorrelated \cite{milner2014a}, $\rho_{\rm tot} = \rho_{\rm cm} \otimes \rho$. In addition, the rotational energy is approximately conserved during a collision if the rotor revolves rapidly enough (see Fig.~\ref{fig:fig2} and App.~\ref{app:b}), $f_{jm,j'm'}({\bf q},{\bf q}') = \delta_{jj'} f_{jm,jm'}({\bf q},{\bf q}')$. Tracing out the center-of -mass under these two simplifications yields after a straight-forward but lengthy calculation the master equation for the rotation state $\rho$,
\begin{equation}
\partial_t \rho = - \frac{i}{\hbar} \left [ \oH + \widetilde \oH_n, \rho \right ] + \widetilde \cL \rho,
\end{equation}
with
\begin{equation}
\widetilde \oH_n = -2 \pi \hbar^2 \frac{n_{\rm g}}{\mu} \int d^3 {\bf p} d^3 {\bf P} \matel{{\bf P}}{\rho_{\rm cm}}{{\bf P}} \mu_{\rm g}({\bf p}) {\rm Re}\left [\of \left ( {\rm rel}({\bf p},{\bf P}),{\rm rel}({\bf p},{\bf P}) \right ) \right ],
\end{equation}
and
\begin{eqnarray}
\widetilde \cL \rho &= & \frac{n_{\rm g} m_{\rm g}^3}{\mu^4} \int d^3 {\bf P} \int_0^\infty dq q^3  \int_{S_2} d^2 {\bf n} d^2 {\bf n}' \matel{{\bf P}}{\rho_{\rm cm}}{{\bf P}} \mu_{\rm g} \left ( \frac{m_{\rm g}}{\mu} q {\bf n}' + \frac{m_{\rm g}}{M} {\bf P}\right ) \notag \\
&& \times \left [ \of ( q {\bf n},q {\bf n}') \rho \of^\dagger ( q {\bf n},q {\bf n}') - \frac{1}{2} \left \{\of^\dagger ( q {\bf n},q {\bf n}')\of ( q {\bf n},q {\bf n}'),\rho \right \} \right ].
\end{eqnarray}

Finally, using that the center-of-mass momentum ${\bf P}$ is thermalized with the gas and integrating it out yields the master equation presented in the main text of the manuscript.

\section{Scattering Amplitude} \label{app:b}

In order to calculate the rotation-state dependent scattering amplitude $f_{jm,j'm'}(q{\bf n},q{\bf n}')$ we start with the time-dependent Schr\"odinger equation in relative coordinates,
\begin{equation} \label{eq:sct}
i \hbar \partial_t \ket{\Psi} = \left [ \frac{\oq^2}{2 \mu} + \frac{\oJ^2}{2 I} + V(\orcm, \om) \right ] \ket{\Psi},
\end{equation}
where $V({\bf r}, {\bf m})$ is the attractive van-der-Waals potential (see main text) and $\oq$ is the relative momentum operator. (Operators are denoted by sans-serif characters.) Expanding the total state as
\begin{equation}
\ket{\Psi} = \sum_{j = 0}^\infty e^{- i E_j t/\hbar} \sum_{m = -j}^j \ket{\chi_{jm}} \ket{jm},
\end{equation}
yields the coupled Schr\"odinger equations
\begin{equation} \label{eq:b3}
i \hbar\partial_t \ket{\chi_{jm}} = \frac{\oq^2}{2 \mu} \ket{\chi_{jm}} + \sum_{j' = 0}^{\infty} e^{i (E_j - E_{j'})t/\hbar} \sum_{m' = -j'}^{j'} \matel{jm}{V(\orcm,\om)}{j'm'} \ket{\chi_{j'm'}}.
\end{equation}

If the molecule rotates multiple times during the collision, $E_j t/\hbar \gg 1$, the exponential function in \eqref{eq:b3} averages to zero for $j \neq j'$ and the different $j$ values are effectively decoupled. In this limit, one obtains for each $j$ a vectorial Schr\"odinger equation for the vector $\left (\ket{\underline{\chi}_j} \right )_m = \ket{\chi_{jm}}$ 
\begin{equation}\label{eq:schr}
i \hbar \partial_t \ket{\underline{\chi}_j} = \left [\frac{\oq^2}{2 \mu} + \rV_j(\orcm) \right ]\ket{\underline{\chi}_j}
\end{equation}
involving the coupling matrix $(\rV_j)_{mm'}(\orcm) = \matel{jm}{V(\orcm,{\bf m})}{j'm'}$.

The vectorial scattering problem described by Eq.~\eqref{eq:schr} can be solved in the eikonal approximation \cite{sakurai} (Schiff's approximation \cite{schiff1956}) for incoming relative momentum $q {\bf n}'$ and outgoing relative momentum $q {\bf n}$, yielding the (matrix-valued) scattering amplitude
\begin{equation} \label{eq:schiff}
\rf_j(q {\bf n},q{\bf  n}') = - i \frac{q}{2 \pi \hbar} \int_{{\bf b} \bot {\bf n'}} d^2 {\bf b} e^{- i q {\bf b} \cdot {\bf n}/ \hbar}\left [ \exp \left \{ - \frac{i \mu}{\hbar q}  \int_{-\infty}^\infty d z \rV_j \left ( {\bf b} + z {\bf n}' \right ) \right \} - \un_j \right ].
\end{equation}
Here $\un_j$ denotes the $(2j+1)\times(2j+1)$ dimensional unity matrix. The desired state-dependent scattering amplitudes are $f_{jm,jm'}(q {\bf n},q{\bf  n}')  = (\rf_j)_{mm'}(q {\bf n},q{\bf  n}') $. They can be explicitly calculated by carrying out the integrations in \eqref{eq:schiff}.

In particular, we first integrate the coupling matrix along the eikonal trajectory $z$ in the exponent. Since the resulting matrix is hermitian, the spectral theorem can be applied to rewrite the matrix exponential in terms of trigonometric functions, so that the radial integration over the impact parameter $b$ can be carried out. We now exploit that forward scattering gives the strongest contribution to the scattering amplitude in order to limit ourselves to calculating $\rf_j(q {\bf n}',q{\bf  n}')$. Making use of the integrals
\begin{subequations}
\begin{equation}
\int_0^\infty db b \sin \left ( \frac{a}{b^5} \right ) = \frac{1}{2}{\rm sign}(a) |a|^{2/5} \Gamma \left (\frac{3}{5} \right ) \cos \left ( \frac{3 \pi}{10} \right )
\end{equation}
and
\begin{equation}
\int_0^\infty db b \sin^2 \left ( \frac{a}{2b^5} \right ) = \frac{1}{4} |a|^{2/5} \Gamma \left (\frac{3}{5} \right ) \sin \left ( \frac{3 \pi}{10} \right )
\end{equation}
\end{subequations}
yields
\begin{equation} \label{eq:fwd}
\rf_j(q {\bf n}',q{\bf  n}') = \frac{q^{3/5}}{4 \pi \hbar} \Gamma \left ( \frac{3}{5} \right ) \left ( \frac{3 \pi \mu C_6}{8 \hbar} \right )^{2/5} e^{i 3 \pi / 10}\oint_{{\bf e}_b \bot {\bf n}'} d {\bf e}_b \left [ \un_j + \rB_j({\bf n}',{\bf e}_b) \right ]^{2/5},
\end{equation}
where ${\bf e}_b = {\bf b}/b$ and the only nonzero entries of the tridiagonal matrix $\rB_j$ are given by
\begin{subequations}
\begin{eqnarray}
(\rB_j)_{mm}({\bf n}',{\bf e}_b) & = & - \frac{\Delta \alpha}{3 \overline{\alpha}} \sqrt{\frac{j(j+1)}{(2 j -1)(2 j +3)}} P_2 \left ( \frac{2 m}{2 j+1} \right ) \left [ \frac{5}{2} ({\bf e}_b \cdot {\bf e}_z)^2 + \frac{1}{2} ({\bf n}' \cdot {\bf e}_z)^2 - 1 \right ] \\
(\rB_j)_{mm\pm1}({\bf n}',{\bf e}_b) & = &  \frac{\Delta \alpha}{18 \overline{\alpha}} \sqrt{\frac{j(j+1)}{(2 j -1)(2 j +3)}} P^1_2 \left ( \frac{2 m \pm 1}{2 j+1} \right ) \left [ 5 {\bf e}_b \cdot {\bf e}_z ({\bf e}_x \pm i {\bf e}_y) \cdot {\bf e}_b +  {\bf n}'\cdot {\bf e}_z ({\bf e}_x \pm i {\bf e}_y) \cdot {\bf n}' \right ] \\
(\rB_j)_{mm\pm 2}({\bf n}',{\bf e}_b) & = &  -\frac{\Delta \alpha}{72 \overline{\alpha}} \sqrt{\frac{j(j+1)}{(2 j -1)(2 j +3)}} P^2_2 \left ( \frac{2 m \pm 2}{2 j+1} \right ) \left [ 5 [{\bf e}_b \cdot ({\bf e}_x \pm i {\bf e}_y)]^2 + [ {\bf n}' \cdot ({\bf e}_x \pm i {\bf e}_y)]^2 \right ].
\end{eqnarray}
\end{subequations}

In order to carry out the ${\bf e}_b$ integration in \eqref{eq:fwd}, we note that 
\begin{equation}
\left [ \un_j + \rB_j({\bf n}',{\bf e}_b) \right ]^{2/5} \simeq \un_j + \frac{2}{5} \rB_j({\bf n}',{\bf e}_b),
\end{equation}
and use the integrals
\begin{subequations}
\begin{eqnarray}
\frac{1}{2 \pi} \oint_{{\bf e}_b \bot {\bf n}'} d {\bf e}_b~ ({\bf e}_b \cdot {\bf e}_z)^2 = \frac{1}{2} \vert {\bf n}' \times {\bf e}_z \vert^2 \\
\frac{1}{2 \pi} \oint_{{\bf e}_b \bot {\bf n}'} d {\bf e}_b ~[{\bf e}_b \cdot {\bf e}_z ({\bf e}_x \pm i {\bf e}_y) \cdot {\bf e}_b]= -\frac{1}{2} {\bf n}' \cdot {\bf e}_z ({\bf e}_x \pm i {\bf e}_y)\cdot {\bf n}' \\
\frac{1}{2 \pi} \oint_{{\bf e}_b \bot {\bf n}'} d {\bf e}_b ~[ {\bf e}_b \cdot ({\bf e}_x \pm i {\bf e}_y) ]^2= -\frac{1}{2} [ {\bf n}' \cdot ({\bf e}_x \pm i {\bf e}_y)]^2.
\end{eqnarray}
\end{subequations}
Inserting this into \eqref{eq:fwd} yields the forward scattering amplitudes. The latter can be inserted into the general form of the decay rate, yielding after integration over ${\bf n}'$ and ${\bf n}$ (giving a factor of $2 \pi$ because the amplitude is strongly peaked in the forward direction) the expression in the main text.
\end{widetext}

\begin{thebibliography}{37}%
\makeatletter
\providecommand \@ifxundefined [1]{%
 \@ifx{#1\undefined}
}%
\providecommand \@ifnum [1]{%
 \ifnum #1\expandafter \@firstoftwo
 \else \expandafter \@secondoftwo
 \fi
}%
\providecommand \@ifx [1]{%
 \ifx #1\expandafter \@firstoftwo
 \else \expandafter \@secondoftwo
 \fi
}%
\providecommand \natexlab [1]{#1}%
\providecommand \enquote  [1]{``#1''}%
\providecommand \bibnamefont  [1]{#1}%
\providecommand \bibfnamefont [1]{#1}%
\providecommand \citenamefont [1]{#1}%
\providecommand \href@noop [0]{\@secondoftwo}%
\providecommand \href [0]{\begingroup \@sanitize@url \@href}%
\providecommand \@href[1]{\@@startlink{#1}\@@href}%
\providecommand \@@href[1]{\endgroup#1\@@endlink}%
\providecommand \@sanitize@url [0]{\catcode `\\12\catcode `\$12\catcode
  `\&12\catcode `\#12\catcode `\^12\catcode `\_12\catcode `\%12\relax}%
\providecommand \@@startlink[1]{}%
\providecommand \@@endlink[0]{}%
\providecommand \url  [0]{\begingroup\@sanitize@url \@url }%
\providecommand \@url [1]{\endgroup\@href {#1}{\urlprefix }}%
\providecommand \urlprefix  [0]{URL }%
\providecommand \Eprint [0]{\href }%
\providecommand \doibase [0]{http://dx.doi.org/}%
\providecommand \selectlanguage [0]{\@gobble}%
\providecommand \bibinfo  [0]{\@secondoftwo}%
\providecommand \bibfield  [0]{\@secondoftwo}%
\providecommand \translation [1]{[#1]}%
\providecommand \BibitemOpen [0]{}%
\providecommand \bibitemStop [0]{}%
\providecommand \bibitemNoStop [0]{.\EOS\space}%
\providecommand \EOS [0]{\spacefactor3000\relax}%
\providecommand \BibitemShut  [1]{\csname bibitem#1\endcsname}%
\let\auto@bib@innerbib\@empty
\bibitem [{\citenamefont {Stapelfeldt}\ and\ \citenamefont
  {Seideman}(2003)}]{stapelfeldt2003}%
  \BibitemOpen
  \bibfield  {author} {\bibinfo {author} {\bibfnamefont {H.}~\bibnamefont
  {Stapelfeldt}}\ and\ \bibinfo {author} {\bibfnamefont {T.}~\bibnamefont
  {Seideman}},\ }\href {\doibase 10.1103/RevModPhys.75.543} {\bibfield
  {journal} {\bibinfo  {journal} {Rev. Mod. Phys.}\ }\textbf {\bibinfo {volume}
  {75}},\ \bibinfo {pages} {543} (\bibinfo {year} {2003})}\BibitemShut
  {NoStop}%
\bibitem [{\citenamefont {Lemeshko}\ \emph {et~al.}(2013)\citenamefont
  {Lemeshko}, \citenamefont {Krems}, \citenamefont {Doyle},\ and\ \citenamefont
  {Kais}}]{lemeshko2013}%
  \BibitemOpen
  \bibfield  {author} {\bibinfo {author} {\bibfnamefont {M.}~\bibnamefont
  {Lemeshko}}, \bibinfo {author} {\bibfnamefont {R.~V.}\ \bibnamefont {Krems}},
  \bibinfo {author} {\bibfnamefont {J.~M.}\ \bibnamefont {Doyle}}, \ and\
  \bibinfo {author} {\bibfnamefont {S.}~\bibnamefont {Kais}},\ }\href@noop {}
  {\bibfield  {journal} {\bibinfo  {journal} {Mol. Phys.}\ }\textbf {\bibinfo
  {volume} {111}},\ \bibinfo {pages} {1648} (\bibinfo {year}
  {2013})}\BibitemShut {NoStop}%
\bibitem [{\citenamefont {Moses}\ \emph {et~al.}(2017)\citenamefont {Moses},
  \citenamefont {Covey}, \citenamefont {Miecnikowski}, \citenamefont {Jin},\
  and\ \citenamefont {Ye}}]{moses2017}%
  \BibitemOpen
  \bibfield  {author} {\bibinfo {author} {\bibfnamefont {S.~A.}\ \bibnamefont
  {Moses}}, \bibinfo {author} {\bibfnamefont {J.~P.}\ \bibnamefont {Covey}},
  \bibinfo {author} {\bibfnamefont {M.~T.}\ \bibnamefont {Miecnikowski}},
  \bibinfo {author} {\bibfnamefont {D.~S.}\ \bibnamefont {Jin}}, \ and\
  \bibinfo {author} {\bibfnamefont {J.}~\bibnamefont {Ye}},\ }\href@noop {}
  {\bibfield  {journal} {\bibinfo  {journal} {Nat. Phys.}\ }\textbf {\bibinfo
  {volume} {13}},\ \bibinfo {pages} {13} (\bibinfo {year} {2017})}\BibitemShut
  {NoStop}%
\bibitem [{\citenamefont {Ohshima}\ and\ \citenamefont
  {Hasegawa}(2010)}]{ohshima2010}%
  \BibitemOpen
  \bibfield  {author} {\bibinfo {author} {\bibfnamefont {Y.}~\bibnamefont
  {Ohshima}}\ and\ \bibinfo {author} {\bibfnamefont {H.}~\bibnamefont
  {Hasegawa}},\ }\href@noop {} {\bibfield  {journal} {\bibinfo  {journal} {Int.
  Rev. Phys. Chem.}\ }\textbf {\bibinfo {volume} {29}},\ \bibinfo {pages} {619}
  (\bibinfo {year} {2010})}\BibitemShut {NoStop}%
\bibitem [{\citenamefont {Fleischer}\ \emph {et~al.}(2012)\citenamefont
  {Fleischer}, \citenamefont {Khodorkovsky}, \citenamefont {Gershnabel},
  \citenamefont {Prior},\ and\ \citenamefont {Averbukh}}]{fleischer2012}%
  \BibitemOpen
  \bibfield  {author} {\bibinfo {author} {\bibfnamefont {S.}~\bibnamefont
  {Fleischer}}, \bibinfo {author} {\bibfnamefont {Y.}~\bibnamefont
  {Khodorkovsky}}, \bibinfo {author} {\bibfnamefont {E.}~\bibnamefont
  {Gershnabel}}, \bibinfo {author} {\bibfnamefont {Y.}~\bibnamefont {Prior}}, \
  and\ \bibinfo {author} {\bibfnamefont {I.~S.}\ \bibnamefont {Averbukh}},\
  }\href@noop {} {\bibfield  {journal} {\bibinfo  {journal} {Isr. J. Chem.}\
  }\textbf {\bibinfo {volume} {52}},\ \bibinfo {pages} {414} (\bibinfo {year}
  {2012})}\BibitemShut {NoStop}%
\bibitem [{\citenamefont {Karczmarek}\ \emph {et~al.}(1999)\citenamefont
  {Karczmarek}, \citenamefont {Wright}, \citenamefont {Corkum},\ and\
  \citenamefont {Ivanov}}]{karczmarek1999}%
  \BibitemOpen
  \bibfield  {author} {\bibinfo {author} {\bibfnamefont {J.}~\bibnamefont
  {Karczmarek}}, \bibinfo {author} {\bibfnamefont {J.}~\bibnamefont {Wright}},
  \bibinfo {author} {\bibfnamefont {P.}~\bibnamefont {Corkum}}, \ and\ \bibinfo
  {author} {\bibfnamefont {M.}~\bibnamefont {Ivanov}},\ }\href {\doibase
  10.1103/PhysRevLett.82.3420} {\bibfield  {journal} {\bibinfo  {journal}
  {Phys. Rev. Lett.}\ }\textbf {\bibinfo {volume} {82}},\ \bibinfo {pages}
  {3420} (\bibinfo {year} {1999})}\BibitemShut {NoStop}%
\bibitem [{\citenamefont {Villeneuve}\ \emph {et~al.}(2000)\citenamefont
  {Villeneuve}, \citenamefont {Aseyev}, \citenamefont {Dietrich}, \citenamefont
  {Spanner}, \citenamefont {Ivanov},\ and\ \citenamefont
  {Corkum}}]{villeneuve2000}%
  \BibitemOpen
  \bibfield  {author} {\bibinfo {author} {\bibfnamefont {D.~M.}\ \bibnamefont
  {Villeneuve}}, \bibinfo {author} {\bibfnamefont {S.~A.}\ \bibnamefont
  {Aseyev}}, \bibinfo {author} {\bibfnamefont {P.}~\bibnamefont {Dietrich}},
  \bibinfo {author} {\bibfnamefont {M.}~\bibnamefont {Spanner}}, \bibinfo
  {author} {\bibfnamefont {M.~Y.}\ \bibnamefont {Ivanov}}, \ and\ \bibinfo
  {author} {\bibfnamefont {P.~B.}\ \bibnamefont {Corkum}},\ }\href {\doibase
  10.1103/PhysRevLett.85.542} {\bibfield  {journal} {\bibinfo  {journal} {Phys.
  Rev. Lett.}\ }\textbf {\bibinfo {volume} {85}},\ \bibinfo {pages} {542}
  (\bibinfo {year} {2000})}\BibitemShut {NoStop}%
\bibitem [{\citenamefont {Yuan}\ \emph {et~al.}(2011)\citenamefont {Yuan},
  \citenamefont {Teitelbaum}, \citenamefont {Robinson},\ and\ \citenamefont
  {Mullin}}]{yuan2011}%
  \BibitemOpen
  \bibfield  {author} {\bibinfo {author} {\bibfnamefont {L.}~\bibnamefont
  {Yuan}}, \bibinfo {author} {\bibfnamefont {S.~W.}\ \bibnamefont
  {Teitelbaum}}, \bibinfo {author} {\bibfnamefont {A.}~\bibnamefont
  {Robinson}}, \ and\ \bibinfo {author} {\bibfnamefont {A.~S.}\ \bibnamefont
  {Mullin}},\ }\href@noop {} {\bibfield  {journal} {\bibinfo  {journal} {Proc.
  Nat. Acad. Sci. USA}\ }\textbf {\bibinfo {volume} {108}},\ \bibinfo {pages}
  {6872} (\bibinfo {year} {2011})}\BibitemShut {NoStop}%
\bibitem [{\citenamefont {Toro}\ \emph {et~al.}(2013)\citenamefont {Toro},
  \citenamefont {Liu}, \citenamefont {Echebiri},\ and\ \citenamefont
  {Mullin}}]{toro2013}%
  \BibitemOpen
  \bibfield  {author} {\bibinfo {author} {\bibfnamefont {C.}~\bibnamefont
  {Toro}}, \bibinfo {author} {\bibfnamefont {Q.}~\bibnamefont {Liu}}, \bibinfo
  {author} {\bibfnamefont {G.~O.}\ \bibnamefont {Echebiri}}, \ and\ \bibinfo
  {author} {\bibfnamefont {A.~S.}\ \bibnamefont {Mullin}},\ }\href@noop {}
  {\bibfield  {journal} {\bibinfo  {journal} {Mol. Phys.}\ }\textbf {\bibinfo
  {volume} {111}},\ \bibinfo {pages} {1892} (\bibinfo {year}
  {2013})}\BibitemShut {NoStop}%
\bibitem [{\citenamefont {Milner}\ \emph
  {et~al.}(2014{\natexlab{a}})\citenamefont {Milner}, \citenamefont
  {Korobenko}, \citenamefont {Hepburn},\ and\ \citenamefont
  {Milner}}]{milner2014a}%
  \BibitemOpen
  \bibfield  {author} {\bibinfo {author} {\bibfnamefont {A.~A.}\ \bibnamefont
  {Milner}}, \bibinfo {author} {\bibfnamefont {A.}~\bibnamefont {Korobenko}},
  \bibinfo {author} {\bibfnamefont {J.~W.}\ \bibnamefont {Hepburn}}, \ and\
  \bibinfo {author} {\bibfnamefont {V.}~\bibnamefont {Milner}},\ }\href@noop {}
  {\bibfield  {journal} {\bibinfo  {journal} {Phys. Rev. Lett.}\ }\textbf
  {\bibinfo {volume} {113}},\ \bibinfo {pages} {043005} (\bibinfo {year}
  {2014}{\natexlab{a}})}\BibitemShut {NoStop}%
\bibitem [{\citenamefont {Korobenko}\ \emph {et~al.}(2014)\citenamefont
  {Korobenko}, \citenamefont {Milner},\ and\ \citenamefont
  {Milner}}]{korobenko2014}%
  \BibitemOpen
  \bibfield  {author} {\bibinfo {author} {\bibfnamefont {A.}~\bibnamefont
  {Korobenko}}, \bibinfo {author} {\bibfnamefont {A.~A.}\ \bibnamefont
  {Milner}}, \ and\ \bibinfo {author} {\bibfnamefont {V.}~\bibnamefont
  {Milner}},\ }\href {\doibase 10.1103/PhysRevLett.112.113004} {\bibfield
  {journal} {\bibinfo  {journal} {Phys. Rev. Lett.}\ }\textbf {\bibinfo
  {volume} {112}},\ \bibinfo {pages} {113004} (\bibinfo {year}
  {2014})}\BibitemShut {NoStop}%
\bibitem [{\citenamefont {Milner}\ \emph
  {et~al.}(2015{\natexlab{a}})\citenamefont {Milner}, \citenamefont
  {Korobenko}, \citenamefont {Flo\ss{}}, \citenamefont {Averbukh},\ and\
  \citenamefont {Milner}}]{milner2015}%
  \BibitemOpen
  \bibfield  {author} {\bibinfo {author} {\bibfnamefont {A.~A.}\ \bibnamefont
  {Milner}}, \bibinfo {author} {\bibfnamefont {A.}~\bibnamefont {Korobenko}},
  \bibinfo {author} {\bibfnamefont {J.}~\bibnamefont {Flo\ss{}}}, \bibinfo
  {author} {\bibfnamefont {I.~S.}\ \bibnamefont {Averbukh}}, \ and\ \bibinfo
  {author} {\bibfnamefont {V.}~\bibnamefont {Milner}},\ }\href {\doibase
  10.1103/PhysRevLett.115.033005} {\bibfield  {journal} {\bibinfo  {journal}
  {Phys. Rev. Lett.}\ }\textbf {\bibinfo {volume} {115}},\ \bibinfo {pages}
  {033005} (\bibinfo {year} {2015}{\natexlab{a}})}\BibitemShut {NoStop}%
\bibitem [{\citenamefont {Milner}\ \emph {et~al.}(2017)\citenamefont {Milner},
  \citenamefont {Korobenko},\ and\ \citenamefont {Milner}}]{milner2017}%
  \BibitemOpen
  \bibfield  {author} {\bibinfo {author} {\bibfnamefont {A.~A.}\ \bibnamefont
  {Milner}}, \bibinfo {author} {\bibfnamefont {A.}~\bibnamefont {Korobenko}}, \
  and\ \bibinfo {author} {\bibfnamefont {V.}~\bibnamefont {Milner}},\
  }\href@noop {} {\bibfield  {journal} {\bibinfo  {journal} {Phys. Rev. Lett.}\
  }\textbf {\bibinfo {volume} {118}},\ \bibinfo {pages} {243201} (\bibinfo
  {year} {2017})}\BibitemShut {NoStop}%
\bibitem [{\citenamefont {Li}\ \emph {et~al.}(2000)\citenamefont {Li},
  \citenamefont {Bahns},\ and\ \citenamefont {Stwalley}}]{li2000}%
  \BibitemOpen
  \bibfield  {author} {\bibinfo {author} {\bibfnamefont {J.}~\bibnamefont
  {Li}}, \bibinfo {author} {\bibfnamefont {J.~T.}\ \bibnamefont {Bahns}}, \
  and\ \bibinfo {author} {\bibfnamefont {W.~C.}\ \bibnamefont {Stwalley}},\
  }\href@noop {} {\bibfield  {journal} {\bibinfo  {journal} {J. Chem. Phys.}\
  }\textbf {\bibinfo {volume} {112}},\ \bibinfo {pages} {6255} (\bibinfo {year}
  {2000})}\BibitemShut {NoStop}%
\bibitem [{\citenamefont {Hartmann}\ and\ \citenamefont
  {Boulet}(2012)}]{hartmann2012}%
  \BibitemOpen
  \bibfield  {author} {\bibinfo {author} {\bibfnamefont {J.-M.}\ \bibnamefont
  {Hartmann}}\ and\ \bibinfo {author} {\bibfnamefont {C.}~\bibnamefont
  {Boulet}},\ }\href@noop {} {\bibfield  {journal} {\bibinfo  {journal} {J.
  Chem. Phys.}\ }\textbf {\bibinfo {volume} {136}},\ \bibinfo {pages} {184302}
  (\bibinfo {year} {2012})}\BibitemShut {NoStop}%
\bibitem [{\citenamefont {Forrey}(2001)}]{forrey2001}%
  \BibitemOpen
  \bibfield  {author} {\bibinfo {author} {\bibfnamefont {R.~C.}\ \bibnamefont
  {Forrey}},\ }\href {\doibase 10.1103/PhysRevA.63.051403} {\bibfield
  {journal} {\bibinfo  {journal} {Phys. Rev. A}\ }\textbf {\bibinfo {volume}
  {63}},\ \bibinfo {pages} {051403} (\bibinfo {year} {2001})}\BibitemShut
  {NoStop}%
\bibitem [{\citenamefont {al~Qady}\ \emph {et~al.}(2011)\citenamefont
  {al~Qady}, \citenamefont {Forrey}, \citenamefont {Yang}, \citenamefont
  {Stancil},\ and\ \citenamefont {Balakrishnan}}]{alqady2011}%
  \BibitemOpen
  \bibfield  {author} {\bibinfo {author} {\bibfnamefont {W.~H.}\ \bibnamefont
  {al~Qady}}, \bibinfo {author} {\bibfnamefont {R.~C.}\ \bibnamefont {Forrey}},
  \bibinfo {author} {\bibfnamefont {B.~H.}\ \bibnamefont {Yang}}, \bibinfo
  {author} {\bibfnamefont {P.~C.}\ \bibnamefont {Stancil}}, \ and\ \bibinfo
  {author} {\bibfnamefont {N.}~\bibnamefont {Balakrishnan}},\ }\href {\doibase
  10.1103/PhysRevA.84.054701} {\bibfield  {journal} {\bibinfo  {journal} {Phys.
  Rev. A}\ }\textbf {\bibinfo {volume} {84}},\ \bibinfo {pages} {054701}
  (\bibinfo {year} {2011})}\BibitemShut {NoStop}%
\bibitem [{\citenamefont {Milner}\ \emph
  {et~al.}(2014{\natexlab{b}})\citenamefont {Milner}, \citenamefont
  {Korobenko},\ and\ \citenamefont {Milner}}]{milner2014b}%
  \BibitemOpen
  \bibfield  {author} {\bibinfo {author} {\bibfnamefont {A.~A.}\ \bibnamefont
  {Milner}}, \bibinfo {author} {\bibfnamefont {A.}~\bibnamefont {Korobenko}}, \
  and\ \bibinfo {author} {\bibfnamefont {V.}~\bibnamefont {Milner}},\
  }\href@noop {} {\bibfield  {journal} {\bibinfo  {journal} {New J. Phys.}\
  }\textbf {\bibinfo {volume} {16}},\ \bibinfo {pages} {093038} (\bibinfo
  {year} {2014}{\natexlab{b}})}\BibitemShut {NoStop}%
\bibitem [{\citenamefont {Khodorkovsky}\ \emph {et~al.}(2015)\citenamefont
  {Khodorkovsky}, \citenamefont {Steinitz}, \citenamefont {Hartmann},\ and\
  \citenamefont {Averbukh}}]{khodorkovsky2015}%
  \BibitemOpen
  \bibfield  {author} {\bibinfo {author} {\bibfnamefont {Y.}~\bibnamefont
  {Khodorkovsky}}, \bibinfo {author} {\bibfnamefont {U.}~\bibnamefont
  {Steinitz}}, \bibinfo {author} {\bibfnamefont {J.-M.}\ \bibnamefont
  {Hartmann}}, \ and\ \bibinfo {author} {\bibfnamefont {I.~S.}\ \bibnamefont
  {Averbukh}},\ }\href@noop {} {\bibfield  {journal} {\bibinfo  {journal} {Nat.
  Commun.}\ }\textbf {\bibinfo {volume} {6}},\ \bibinfo {pages} {7791}
  (\bibinfo {year} {2015})}\BibitemShut {NoStop}%
\bibitem [{\citenamefont {Milner}\ \emph
  {et~al.}(2015{\natexlab{b}})\citenamefont {Milner}, \citenamefont
  {Korobenko}, \citenamefont {Rezaiezadeh},\ and\ \citenamefont
  {Milner}}]{milner2015b}%
  \BibitemOpen
  \bibfield  {author} {\bibinfo {author} {\bibfnamefont {A.~A.}\ \bibnamefont
  {Milner}}, \bibinfo {author} {\bibfnamefont {A.}~\bibnamefont {Korobenko}},
  \bibinfo {author} {\bibfnamefont {K.}~\bibnamefont {Rezaiezadeh}}, \ and\
  \bibinfo {author} {\bibfnamefont {V.}~\bibnamefont {Milner}},\ }\href
  {\doibase 10.1103/PhysRevX.5.031041} {\bibfield  {journal} {\bibinfo
  {journal} {Phys. Rev. X}\ }\textbf {\bibinfo {volume} {5}},\ \bibinfo {pages}
  {031041} (\bibinfo {year} {2015}{\natexlab{b}})}\BibitemShut {NoStop}%
\bibitem [{\citenamefont {Hornberger}(2007)}]{hornberger2007}%
  \BibitemOpen
  \bibfield  {author} {\bibinfo {author} {\bibfnamefont {K.}~\bibnamefont
  {Hornberger}},\ }\href@noop {} {\bibfield  {journal} {\bibinfo  {journal}
  {Europhys. Lett.}\ }\textbf {\bibinfo {volume} {77}},\ \bibinfo {pages}
  {50007} (\bibinfo {year} {2007})}\BibitemShut {NoStop}%
\bibitem [{\citenamefont {Hornberger}\ and\ \citenamefont
  {Vacchini}(2008)}]{hornberger2008}%
  \BibitemOpen
  \bibfield  {author} {\bibinfo {author} {\bibfnamefont {K.}~\bibnamefont
  {Hornberger}}\ and\ \bibinfo {author} {\bibfnamefont {B.}~\bibnamefont
  {Vacchini}},\ }\href@noop {} {\bibfield  {journal} {\bibinfo  {journal}
  {Phys. Rev. A}\ }\textbf {\bibinfo {volume} {77}},\ \bibinfo {pages} {022112}
  (\bibinfo {year} {2008})}\BibitemShut {NoStop}%
\bibitem [{\citenamefont {Smirne}\ and\ \citenamefont
  {Vacchini}(2010)}]{smirne2010}%
  \BibitemOpen
  \bibfield  {author} {\bibinfo {author} {\bibfnamefont {A.}~\bibnamefont
  {Smirne}}\ and\ \bibinfo {author} {\bibfnamefont {B.}~\bibnamefont
  {Vacchini}},\ }\href@noop {} {\bibfield  {journal} {\bibinfo  {journal}
  {Phys. Rev. A}\ }\textbf {\bibinfo {volume} {82}},\ \bibinfo {pages} {042111}
  (\bibinfo {year} {2010})}\BibitemShut {NoStop}%
\bibitem [{\citenamefont {Stickler}\ \emph {et~al.}(2016)\citenamefont
  {Stickler}, \citenamefont {Papendell},\ and\ \citenamefont
  {Hornberger}}]{stickler2016b}%
  \BibitemOpen
  \bibfield  {author} {\bibinfo {author} {\bibfnamefont {B.~A.}\ \bibnamefont
  {Stickler}}, \bibinfo {author} {\bibfnamefont {B.}~\bibnamefont {Papendell}},
  \ and\ \bibinfo {author} {\bibfnamefont {K.}~\bibnamefont {Hornberger}},\
  }\href@noop {} {\bibfield  {journal} {\bibinfo  {journal} {Phys. Rev. A}\
  }\textbf {\bibinfo {volume} {94}},\ \bibinfo {pages} {033828} (\bibinfo
  {year} {2016})}\BibitemShut {NoStop}%
\bibitem [{\citenamefont {Zhong}\ and\ \citenamefont
  {Robicheaux}(2016)}]{zhong2016}%
  \BibitemOpen
  \bibfield  {author} {\bibinfo {author} {\bibfnamefont {C.}~\bibnamefont
  {Zhong}}\ and\ \bibinfo {author} {\bibfnamefont {F.}~\bibnamefont
  {Robicheaux}},\ }\href@noop {} {\bibfield  {journal} {\bibinfo  {journal}
  {Phys. Rev. A}\ }\textbf {\bibinfo {volume} {94}},\ \bibinfo {pages} {052109}
  (\bibinfo {year} {2016})}\BibitemShut {NoStop}%
\bibitem [{\citenamefont {Papendell}\ \emph {et~al.}(2017)\citenamefont
  {Papendell}, \citenamefont {Stickler},\ and\ \citenamefont
  {Hornberger}}]{papendell2017}%
  \BibitemOpen
  \bibfield  {author} {\bibinfo {author} {\bibfnamefont {B.}~\bibnamefont
  {Papendell}}, \bibinfo {author} {\bibfnamefont {B.~A.}\ \bibnamefont
  {Stickler}}, \ and\ \bibinfo {author} {\bibfnamefont {K.}~\bibnamefont
  {Hornberger}},\ }\href {http://stacks.iop.org/1367-2630/19/i=12/a=122001}
  {\bibfield  {journal} {\bibinfo  {journal} {New J. Phys.}\ }\textbf {\bibinfo
  {volume} {19}},\ \bibinfo {pages} {122001} (\bibinfo {year}
  {2017})}\BibitemShut {NoStop}%
\bibitem [{\citenamefont {Stickler}\ \emph {et~al.}(2017)\citenamefont
  {Stickler}, \citenamefont {Schrinski},\ and\ \citenamefont
  {Hornberger}}]{stickler2018a}%
  \BibitemOpen
  \bibfield  {author} {\bibinfo {author} {\bibfnamefont {B.~A.}\ \bibnamefont
  {Stickler}}, \bibinfo {author} {\bibfnamefont {B.}~\bibnamefont {Schrinski}},
  \ and\ \bibinfo {author} {\bibfnamefont {K.}~\bibnamefont {Hornberger}},\
  }\href@noop {} {\bibfield  {journal} {\bibinfo  {journal} {arXiv:1712.05163}\
  } (\bibinfo {year} {2017})}\BibitemShut {NoStop}%
\bibitem [{\citenamefont {DePristo}\ \emph {et~al.}(1979)\citenamefont
  {DePristo}, \citenamefont {Augustin}, \citenamefont {Ramaswamy},\ and\
  \citenamefont {Rabitz}}]{depristo1979}%
  \BibitemOpen
  \bibfield  {author} {\bibinfo {author} {\bibfnamefont {A.~E.}\ \bibnamefont
  {DePristo}}, \bibinfo {author} {\bibfnamefont {S.~D.}\ \bibnamefont
  {Augustin}}, \bibinfo {author} {\bibfnamefont {R.}~\bibnamefont {Ramaswamy}},
  \ and\ \bibinfo {author} {\bibfnamefont {H.}~\bibnamefont {Rabitz}},\
  }\href@noop {} {\bibfield  {journal} {\bibinfo  {journal} {J. Chem. Phys.}\
  }\textbf {\bibinfo {volume} {71}},\ \bibinfo {pages} {850} (\bibinfo {year}
  {1979})}\BibitemShut {NoStop}%
\bibitem [{\citenamefont {Schmidt}\ and\ \citenamefont
  {Lemeshko}(2015)}]{schmidt2015}%
  \BibitemOpen
  \bibfield  {author} {\bibinfo {author} {\bibfnamefont {R.}~\bibnamefont
  {Schmidt}}\ and\ \bibinfo {author} {\bibfnamefont {M.}~\bibnamefont
  {Lemeshko}},\ }\href {\doibase 10.1103/PhysRevLett.114.203001} {\bibfield
  {journal} {\bibinfo  {journal} {Phys. Rev. Lett.}\ }\textbf {\bibinfo
  {volume} {114}},\ \bibinfo {pages} {203001} (\bibinfo {year}
  {2015})}\BibitemShut {NoStop}%
\bibitem [{\citenamefont {Armon}\ and\ \citenamefont
  {Friedland}(2016)}]{armon2016}%
  \BibitemOpen
  \bibfield  {author} {\bibinfo {author} {\bibfnamefont {T.}~\bibnamefont
  {Armon}}\ and\ \bibinfo {author} {\bibfnamefont {L.}~\bibnamefont
  {Friedland}},\ }\href@noop {} {\bibfield  {journal} {\bibinfo  {journal}
  {Phys. Rev. A}\ }\textbf {\bibinfo {volume} {93}},\ \bibinfo {pages} {043406}
  (\bibinfo {year} {2016})}\BibitemShut {NoStop}%
\bibitem [{\citenamefont {Armon}\ and\ \citenamefont
  {Friedland}(2017)}]{armon2017}%
  \BibitemOpen
  \bibfield  {author} {\bibinfo {author} {\bibfnamefont {T.}~\bibnamefont
  {Armon}}\ and\ \bibinfo {author} {\bibfnamefont {L.}~\bibnamefont
  {Friedland}},\ }\href@noop {} {\bibfield  {journal} {\bibinfo  {journal}
  {Phys. Rev. A}\ }\textbf {\bibinfo {volume} {96}},\ \bibinfo {pages} {033411}
  (\bibinfo {year} {2017})}\BibitemShut {NoStop}%
\bibitem [{\citenamefont {Hirschfelder}\ \emph {et~al.}(1964)\citenamefont
  {Hirschfelder}, \citenamefont {Bird},\ and\ \citenamefont
  {Curtiss}}]{hirschfelder1964}%
  \BibitemOpen
  \bibfield  {author} {\bibinfo {author} {\bibfnamefont {J.}~\bibnamefont
  {Hirschfelder}}, \bibinfo {author} {\bibfnamefont {R.~B.}\ \bibnamefont
  {Bird}}, \ and\ \bibinfo {author} {\bibfnamefont {C.~F.}\ \bibnamefont
  {Curtiss}},\ }\href@noop {} {\emph {\bibinfo {title} {Molecular theory of
  gases and liquids}}}\ (\bibinfo  {publisher} {Wiley},\ \bibinfo {year}
  {1964})\BibitemShut {NoStop}%
\bibitem [{\citenamefont {Hutson}\ and\ \citenamefont {Green}(1994)}]{molscat}%
  \BibitemOpen
  \bibfield  {author} {\bibinfo {author} {\bibfnamefont {J.~M.}\ \bibnamefont
  {Hutson}}\ and\ \bibinfo {author} {\bibfnamefont {S.}~\bibnamefont {Green}},\
  }\href@noop {} {\emph {\bibinfo {title} {MOLSCAT Vers. 14}}}\ (\bibinfo
  {publisher} {Collaborative Computational Project No. 6 of the Engineering and
  Physical Sciences Research Council (UK)},\ \bibinfo {year}
  {1994})\BibitemShut {NoStop}%
\bibitem [{\citenamefont {Spelsberg}\ and\ \citenamefont
  {Meyer}(1994)}]{spelsberg1994}%
  \BibitemOpen
  \bibfield  {author} {\bibinfo {author} {\bibfnamefont {D.}~\bibnamefont
  {Spelsberg}}\ and\ \bibinfo {author} {\bibfnamefont {W.}~\bibnamefont
  {Meyer}},\ }\href@noop {} {\bibfield  {journal} {\bibinfo  {journal} {J.
  Chem. Phys.}\ }\textbf {\bibinfo {volume} {101}},\ \bibinfo {pages} {1282}
  (\bibinfo {year} {1994})}\BibitemShut {NoStop}%
\bibitem [{\citenamefont {Mortimer}(1993)}]{mortimer1993}%
  \BibitemOpen
  \bibfield  {author} {\bibinfo {author} {\bibfnamefont {R.}~\bibnamefont
  {Mortimer}},\ }\href@noop {} {\emph {\bibinfo {title} {Physical Chemistry}}}\
  (\bibinfo  {publisher} {Benjamin/Cummings Publishing Company},\ \bibinfo
  {year} {1993})\BibitemShut {NoStop}%
\bibitem [{\citenamefont {Schiff}(1956)}]{schiff1956}%
  \BibitemOpen
  \bibfield  {author} {\bibinfo {author} {\bibfnamefont {L.~I.}\ \bibnamefont
  {Schiff}},\ }\href@noop {} {\bibfield  {journal} {\bibinfo  {journal} {Phys.
  Rev.}\ }\textbf {\bibinfo {volume} {103}},\ \bibinfo {pages} {443} (\bibinfo
  {year} {1956})}\BibitemShut {NoStop}%
\bibitem [{\citenamefont {Sakurai}(1993)}]{sakurai}%
  \BibitemOpen
  \bibfield  {author} {\bibinfo {author} {\bibfnamefont {J.~J.}\ \bibnamefont
  {Sakurai}},\ }\href@noop {} {\emph {\bibinfo {title} {{Modern Quantum
  Mechanics}}}},\ \bibinfo {edition} {{Revised}}\ ed.\ (\bibinfo  {publisher}
  {Addison Wesley},\ \bibinfo {address} {Reading, Massachusetts},\ \bibinfo
  {year} {1993})\BibitemShut {NoStop}%
\end{thebibliography}
\end{document}